\documentclass[12pt]{article}
\usepackage{graphicx}

\newcommand\pubnumber{Article 38 in eConf C1304143}
\newcommand\pubdate{\today}

\def\mpik{\footnote{Max-Planck-Institut f\"ur Kernphysik P.O. Box 103980, D 69029 Heidelberg, Germany}}
\def\gatech{\footnote{now at: School of Physics and Center for Relativistic Astrophysics, Georgia Institute of Technology, Atlanta, Georgia, USA}}
\def\durham{\footnote{University of Durham, Department of Physics, South Road, Durham DH1 3LE, U.K.}}
\def\adelaide{\footnote{School of Chemistry \& Physics, University of Adelaide, Adelaide 5005, Australia}}
\def\hsinchu{\footnote{Institute of Astronomy and Department of Physics, National Tsing Hua University, Hsinchu, Taiwan}}

\def\Title#1{\begin{center} {\Large #1 } \end{center}}
\def\Author#1{\begin{center}{ \sc #1} \end{center}}

\newcommand\pubblock{\rightline{\begin{tabular}{l} \pubnumber\\
         \pubdate  \end{tabular}}}
\newenvironment{Abstract}{\begin{quotation}  }{\end{quotation}}
\newenvironment{Presented}{\begin{quotation} \begin{center}
             PRESENTED AT\end{center}\bigskip
      \begin{center}\begin{large}}{\end{large}\end{center} \end{quotation}}
\def\Acknowledgements{\bigskip  \bigskip \begin{center} \begin{large}
             \bf ACKNOWLEDGEMENTS \end{large}\end{center}}

\def\beq{\begin{equation}}
\def\eeq#1{\label{#1}\end{equation}}
\def\eeqn{\end{equation}}

\def\beqa{\begin{eqnarray}}
\def\eeqa#1{\label{#1}\end{eqnarray}}
\def\eeqan{\end{eqnarray}}

\let\bar=\overbar

\def\Dslash{\not{\hbox{\kern-4pt $D$}}}
\def\dslash{\not{\hbox{\kern-2pt $\del$}}}

\def\msb{{\bar{\ssstyle M \kern -1pt S}}}

\def\degr{\hbox{$^\circ$}}
\def\arcmin{\hbox{$^\prime$}}
\def\arcsec{\hbox{$^{\prime\prime}$}}

\newcommand{\linebreakcell}[2][c]{%
  \begin{tabular}[#1]{@{}c@{}}#2\end{tabular}}

\usepackage{overpic}

\begin{document}
\begin{titlepage}
\pubblock

\vfill
\Title{Searching for TeV emission from GRBs: the status of the H.E.S.S. GRB programme}
\vfill
\Author{D.~Lennarz\mpik$^{,}$\gatech, P.M.~Chadwick\durham, W.~Domainko\footnotemark[1], R.D.~Parsons\footnotemark[1], G.~Rowell\adelaide~and~P.H.T.~Tam\hsinchu~for the H.E.S.S. collaboration}
\vfill
\begin{Abstract}
H.E.S.S. is an array of five Imaging Atmospheric Cherenkov Telescopes (IACTs) located 1800~m above sea level in the Khomas Highland of Namibia and is sensitive to very-high-energy (VHE) gamma rays between tens of GeV to tens of TeV. The very-high background rejection capabilities of IACTs provide excellent sensitivity of H.E.S.S. to GRBs. In this contribution the status of the H.E.S.S. GRB programme, already started in 2003, is reviewed. A highlight is the recent addition of the fifth telescope, which is the world's largest IACT. Its 600 square metre mirror lowers the energy threshold to tens of GeV and provides an effective area that is ten thousands of times larger than \emph{Fermi}-LAT at these energies. The higher performance drive system will reduce the response time to a GRB alert, which will significantly enhance the chances of a H.E.S.S. GRB detection. Recent results on selected GRBs will be shown.
\end{Abstract}
\vfill
\begin{Presented}
Huntsville Gamma Ray Burst Symposium\\
Nashville, Tennessee, USA, 14-18 April 2013
\end{Presented}
\vfill
\end{titlepage}
\def\thefootnote{\fnsymbol{footnote}}
\setcounter{footnote}{0}

\section{Introduction}
The prompt keV to MeV emission from GRBs is in general well described by Band functions \cite{Band:1993eg}. \emph{Fermi}-LAT detects a subset of very energetic bursts at higher energies (HE, above $\sim$20 MeV) with a large variety of spectral properties. As an example, for GRB~080916C the emission is consistent with a Band function from keV to GeV energies \cite{Abdo:2009zza}, while for GRB~100724B an exponential cutoff in the Band function was recently detected \cite{LAT_GRB_catalog}. More intriguing and a challenge for GRB modelling are bursts that exhibit an additional hard power-law component (e.g. GRB 090902B and 090510 \cite{Abdo:2009pg,Ackermann:2010us}). Also these components can have a spectral break (GRB~090926A \cite{Bregeon:2011bu}). As shown recently, all bright or well-observed LAT-detected GRBs show significant deviations from a Band function and require additional spectral components \cite{LAT_GRB_catalog}.

Another interesting result of \emph{Fermi} is that the $>100$~MeV emission of GRBs starts systematically later than the emission at lower energies. It can reach up to 40~s for GRB~090626 while a typical value is a few seconds \cite{LAT_GRB_catalog}. Furthermore, the duration of the LAT emission is longer than the keV equivalent, e.g. reaching up to almost 700~s for GRB~090328 \cite{LAT_GRB_catalog}. Both effects appear to be less prominent at lower energies.

The highest energy photon observed by the LAT so far had an energy of 94~GeV (GRB~130427A). Observations at even higher energies are challenging for the LAT because the effective area stays approximately constant at these energies ($\sim0.6$~$\rm m^2$ at 10~GeV) and the GRB photon flux falls quickly with energy. Such observations can help to understand the emission mechanism at work, e.g. the observation of high-energy photons places a lower limit on the Lorentz boost factor in the GRB jet because the photons have not been attenuated via pair production at the source. Furthermore, the observation of a cut-off could also shed light on the extragalactic background light (EBL), where high-energy photons are attenuated via pair production on the way to Earth. However, the EBL also limits the distance up to where very-high-energy (VHE, $>100$~GeV) emission from GRBs can be detected.

\section{Imaging Atmospheric Cherenkov Telescopes}
IACTs detect VHE gamma rays between hundreds of GeV to tens of TeV using Cherenkov light emitted when the gamma ray is absorbed in the atmosphere in an extensive air shower. Observations are typically taken during darkness when the moon is below the horizon (``dark time''). The typical dark time fraction for an IACT is approximately 20\%. Since IACTs are pointed instruments with a limited field of view of a couple of degrees and GRBs happen randomly in the sky, the telescopes have to be repointed quickly. Given the \emph{Fermi} finding of delayed and extended emission at HE, the VHE emission might shows similar effects. Hence, while the follow-up time may not be decisive, it is still desirable to be on target as quickly as possible.

\subsection{The High Energy Stereoscopic System (H.E.S.S.)}
H.E.S.S. consists of five Imaging Atmospheric Cherenkov Telescopes (IACTs) located 1800~m above sea level in the Khomas Highland of Namibia \cite{Aharonian:2006pe}. The first four telescopes of the H.E.S.S. project (phase I) have been operational since December 2003. They have $100~\rm m^2$ tessellated mirror surface arranged in a Davies-Cotton design with a focal length of 15~m. The telescopes are arranged in a square with 120~m side length with one diagonal oriented north-south. Furthermore, each telescope is equipped with a pixelated camera of 960 photomultiplier tubes (PMTs) with Winston cones in front to improve the light collection efficiency. One pixel subtends approximately 0.16\degr, resulting in a total field of view of 5\degr~in diameter. Air shower are triggered on three different stages \cite{Funk:trigger}: at PMT level, at telescope and at array level (central trigger). The reconstruction is done on events seen by at least two of the four telescopes, allowing stereoscopic image analysis. This results in an angular resolution (68\% containment) of typically $0.1\degr$ and an energy resolution of $\sim15\%$.

A fifth telescope (H.E.S.S. II) was added to the center of the existing array in July 2012. It is the world's largest IACT with over 600~$\rm m^2$ mirror area and a nominal focal length of 36~m. The camera features 2048 PMTs, equivalent to a field of view of approximately 3.2\degr~on the sky. In contrast to the existing array, H.E.S.S. II will not only read out stereoscopic events, but also events triggering the new telescope only. The energy threshold of the experiment is lowered to tens of GeV, thus bridging the energy gap towards \emph{Fermi}-LAT. A higher performance drive system will reduce the response time to a GRB trigger (see further discussion below). Both features will significantly improve the chances of a H.E.S.S. GRB detection.

\section{The H.E.S.S. GRB Programme}
H.E.S.S. observations of GRBs have already started in early 2003, when the array was still under construction. Two GRBs in 2009 (030329 and 030821) were observed with two telescopes only. For the first GRB the central trigger was not yet available and coincidence of events was determined offline using GPS time stamps. Since early 2005 the H.E.S.S. data acquisition system (DAQ) is connected via socket connection to the GRB Coordinates Network\footnote{http://gcn.gscfc.nasa.gov} (GCN). A dedicated programme is running at the telescope site that processes the GCN notices in real time.

Currently, H.E.S.S. triggers on notices with a GRB positional uncertainty $<2.5\degr$ from \emph{Swift}-BAT (using GCN type 61, SWIFT\_BAT\_GRB\_POSITION) and \emph{Fermi}-LAT (using GCN type 121, FERMI\_LAT\_GRB\_POS\_UPD). Unfortunately, the LAT has only provided one trigger so far (GRB~090510) which was not visible in Namibia.  Additionally, for the BAT notices it is checked that the position is incompatible with known sources and a filter is applied to the type of source/trigger found (soln\_status). The bit mask to filter the soln\_status\footnote{see http://gcn.gsfc.nasa.gov/sock\_pkt\_def\_doc.html} is 11X0Y0YX0000, which means e.g. that only such triggers are accepted where a point source was found. Furthermore, X means both bits are accepted and Y means a notice with the high background level (i.e. near the South Atlantic Anomaly) flag set (7$^{th}$ bit) is only filtered in case of a rate trigger (4$^{th}$ bit). Furthermore, notices with a cyclic redundancy check error in one or more of the packets are ignored.

\subsection{Automatic Repointing}
Until April 2011, the observers present at the telescopes were only notified in case of a GRB trigger and observations should have been started immediately, if the trigger is received during dark time with good weather conditions and the GRB position can be observed with a zenith angle smaller than $45\degr$ to ensure a reasonably low energy threshold (``prompt alerts''). If a trigger does not meet the dark time or zenith criteria it is observed once they are met if the observations would not start too long after the trigger (``afterglow alert''). The acceptable delay after the trigger is redshift dependent and is 24~h if $z<0.1$, 12~h if $z<0.3$, 6~h if $z<1$ and 4~h if $z$ is unknown. Therefore, H.E.S.S. is very interested in timely determinations of redshifts. The redshift criteria is introduced because for a nearby burst a lower flux at a later stage might still be detectable due to the effect of EBL absorption. For each trigger per default four observations (nominal length of 28 min) are undertaken. In case the online analysis shows a hint of a signal ($> 3\sigma$) the GRB position is observed further. Observations only start when more than thirty minutes of dark time at low zenith angle are available.

In April 2011, the human-in-the-loop process of observers manually stoping the current observation, scheduling the GRB as a new target and starting a new observation, has been replaced with an automated repointing that automates the necessary steps. This leads to a significant reduction of the transition time to the GRB observation. It has been successfully applied in the observation of GRB~120328A.

\subsection{Fermi Triggers}
The current trigger rate of \emph{Swift} is $\sim90$ GRBs per year. Accounting for the H.E.S.S. criteria of a prompt trigger and weather constraints, this leads to about $2-3$ prompt triggers per year. Since less than 10\% of the GRBs detected by \emph{Fermi}-GBM are also seen by the LAT, it becomes apparent that it might take a considerable time until H.E.S.S. promptly observes a GRB with emission in the LAT energy regime.

This problem can be overcome by increasing the H.E.S.S. trigger rate, e.g.~by also including GBM triggers ($\sim250$ per year). However, the bursts located by GBM typically have a large positional error. A comparison between the final GBM position that is available from a GCN notice and the \emph{Swift}-XRT position (in case of a co-detection of both instruments) reveals that a large fraction of the GBM positions would be outside the field of view of an IACT. Furthermore, the position might change considerably between the flight based position generated on the spacecraft and the ground based position using more sophisticated analysis methods. This makes it necessary to update the observation position of the H.E.S.S. telescopes during the observation especially in cases where a later, more accurate position indicates that the GRB is not within or near the edge of the field of view of the current observation.

\subsection{A New Alert Scheme}
With the advent of H.E.S.S. II, the H.E.S.S. GRB programme is currently undergoing a major technical revision with the aim to reduce the transition time to the GRB observation as much as possible. The main contribution to the transition time is the slewing of the telescopes. This process is optimised by operating H.E.S.S. II in reverse slewing mode, where the telescope's elevation axis is allowed to drive through zenith. This is considerably faster and reduces the mean time to go from a random observation position to a random position on the sky from 61~s to 36~s \cite{Hofverberg:ICRC}. Furthermore, the tracking normally undergoes a process of ``fine positioning'' that increases the tracking accuracy. For GRB alerts, even more so for the ones from GBM with a positional error of a couple of degrees, this is not necessary. Thus, for GRB observations the tracking system reports to be on target as soon as the target is on the edge of the field of view. The optimised H.E.S.S. II drive system will lead to the situation that the new telescope is ready for data taking before the H.E.S.S. I telescopes have arrived. Therefore, data is collected as soon as H.E.S.S. II is on target and the other telescopes will join in once they arrive.

During normal operation, the high voltage of the PMTs is switched off when the telescopes are slewing to a new target to prevent damage due to bright stars in the field of view, which causes an additional delay. For GRB observations, the telescopes will immediately start moving to the new target while the ongoing observation is stopped. The cameras are sent to an internal paused state where they stop taking data, but the high voltage stays on. Once on target, the camera is unpaused and camera pixels that might have been automatically turned off due to over currents are reactivated. Moreover, systems that are optional for the data taking, like e.g.~monitoring systems, are not waited for. This reduces the transition time to basically just the slewing time of the telescopes.

\section{H.E.S.S. GRB Observations}
H.E.S.S. has observed 32 GRB positions until August 2007 and reported the results on 22 GRBs. No evidence for a VHE signal was found \cite{Aharonian:2009}. The to date only completely simultaneous observation of a GRB prompt phase with an IACT was GRB~060602B, where the GRB was by chance in the current field of view \cite{Aharonian:2008nk}. Again, no signal was found, neither in the prompt nor afterglow phase. However, the burst was unusually soft and there is strong indication that it was indeed a type I X-ray burst \cite{GRB060602B_no_GRB}. This suggestion is also based on the Galactic position of $l=1.15^\circ$ and $b= -0.3^\circ$.

Table~\ref{tab:GRB_observations} shows all prompt and afterglow GRB observations of H.E.S.S. between August 2007 and March 2012. GRB~091018 was a prompt alert, but the sky was covered at the beginning of the night so observations could not start immediately. This burst is therefore listed in the afterglow section. H.E.S.S. cannot follow up on all possible alerts due to various reasons. The weather conditions prevented observation of two prompt bursts opportunities (GRB~120327A and GRB~100425A) and five afterglow opportunities (GRB~090306B, GRB~090205, GRB~081112, GRB~080218B and GRB~080207). For some bursts no GCN notice is issued, e.g. if the burst is found in a \emph{Swift} ground analysis, thus GRB~080702B (prompt) and GRB~110206A, GRB~101204A, GRB~100203A, GRB~091117, GRB~080802 (afterglow) were not observed. Technical problems, e.g.~because the internet connection to the site is interrupted, prevented the observation of GRB~080605 (prompt) and GRB~090926B, GRB~090628 (afterglow). The nearby burst GRB~071227 ($z<0.4$) was missed because no shift crew was on site to take data. For H.E.S.S. this burst would have been the nearest prompt burst with known redshift. For the afterglow alerts of GRB~100316D (z=0.059) and GRB~090424 (z=0.544) the redshift information was not available in time to start the observations. GRB~100316D was particularly interesting because the burst was associated with SN~2010bh. Afterglow observations for GRB~081028A were started, but immediately stopped due to bad weather.

\subsection{GRB~100621A}
GRB~100621A is in certain aspects the most interesting GRB observed by H.E.S.S. so far. It had a very bright prompt phase, the brightest X-ray afterglow so far observed by the XRT and is the closest burst observed promptly by the H.E.S.S. telescopes. Due to its relatively small redshift ($z=0.542$) it is located within the VHE gamma-ray horizon. H.E.S.S. received the trigger in Namibia at 03:04:01 UT which is 29~s after the \emph{Swift} satellite trigger. However, human intervention delayed the start of the observations to 03:14:55 UT which is 683~s after the trigger. The end of the dark time due to moon rise allowed only two observations each lasting for 30 minutes.

The GRB is analysed using a point source analysis with the standard H.E.S.S. analysis software applying the reflected background method. Table~\ref{tab:results_significance} shows the preliminary results of the analysis. No significant excess is observed for the total dataset. In order to search for emission on shorter time scales and closer to the satellite trigger a further analysis was done on each observation separately and the events corresponding to the first 300~s of the first observation. Shorter time scales are not possible, because the number of events in the on-region would become too low to estimate the significance. No significant excess is found here either.

\begin{figure}[t]
\centering
 \begin{overpic}[width=0.88\textwidth]{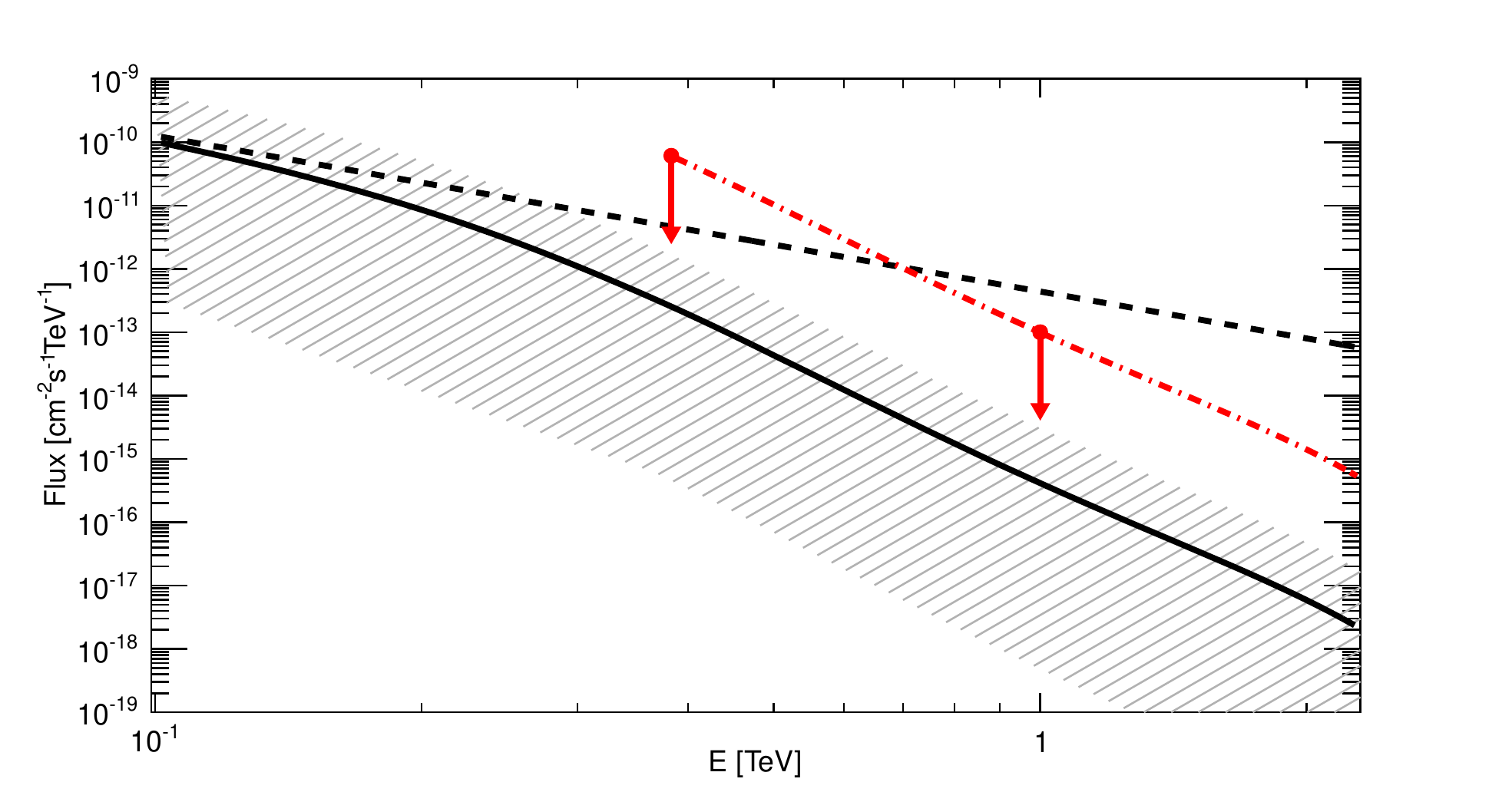}
 \put(15,10){\LARGE Preliminary}
 \end{overpic}
\caption{The solid line shows the extrapolation of the prompt GRB spectrum after correcting for EBL absorption and the delay of the H.E.S.S. observations, while the dashed line shows the unabsorbed power law. The red dashed-dotted line shows the spectrum that corresponds to the H.E.S.S. upper limit. The shaded area shows the extrapolated spectrum if the high-energy photon index is varied within its one-sigma error.}
\label{fig:bandfunction_lightcurve}
\end{figure}

Figure~\ref{fig:bandfunction_lightcurve} compares the H.E.S.S. upper limits to a spectral-temporal model build upon the prompt GRB spectrum. It constrains the possibility that GRB~100621A had an extra hard power-law like observed in other GRBs. Further details and results will be published in a forthcoming publication.

\section{Summary}
Recent results from \emph{Fermi} have shown that a nearby and powerful burst is detectable for IACTs at VHE. The detection will provide a detailed measurement of the spectral cut-off which will amongst other provide input for acceleration theories. With the lower energy threshold of H.E.S.S. II, the automatic repointing and the new alert scheme, leading to a dramatic decrease of the reaction time, H.E.S.S. has very good chances of detection VHE radiation from GRBs in the near future.

\Acknowledgements
\scriptsize{The support of the Namibian authorities and of the University of Namibia in facilitating the construction and operation of H.E.S.S. is gratefully acknowledged, as is the support by the German Ministry for Education and Research (BMBF), the Max Planck Society, the French Ministry for Research, the CNRS-IN2P3 and the Astroparticle Interdisciplinary Programme of the CNRS, the U.K. Science and Technology Facilities Council (STFC), the IPNP of the Charles University, the Czech Science Foundation, the Polish Ministry of Science and Higher Education, the South African Department of Science and Technology and National Research Foundation, and by the University of Namibia. We appreciate the excellent work of the technical support staff in Berlin, Durham, Hamburg, Heidelberg, Palaiseau, Paris, Saclay, and in Namibia in the construction and operation of the equipment.}

\begin{table}[p!]
\begin{center}
\footnotesize{
  \begin{tabular}{c|ccccccc}
    GRB & \linebreakcell[t]{Trigger \\ number} & \linebreakcell[t]{Trigger \\ time [UT]} & \linebreakcell[t]{RAJ2000 \\ DECJ2000} & \linebreakcell[t]{Error \\ $[\arcsec]$} & \linebreakcell[t]{Fluence \\ $[\rm{erg~cm^{-2}}]$} & \linebreakcell[t]{T90 \\ $[s]$} & $z$\\
    \hline
    120328A & 518792 & 03:06:19 & \linebreakcell[t]{$16^{\rm h}06^{\rm m}27.30^{\rm s}$ \\ $-39\degr20\arcmin08.3\arcsec$} & 1.9 & $(4.7\pm0.6)10^{-7}$ & $24.2\pm6.7$ & $\sim3.2$\\

    100621A & 425151 & 03:03:32 & \linebreakcell[t]{$21^{\rm h}01^{\rm m}13.12^{\rm s}$ \\ $-51\degr06\arcmin22.5\arcsec$} & 1.7 & ($2.1\pm0.0)\cdot10^{-5}$ & $63.6\pm1.7$ & 0.542\\

    081230 & 338633 & 20:36:12 & \linebreakcell[t]{$02^{\rm h}29^{\rm m}19.53^{\rm s}$ \\ $-25\degr08\arcmin51.8\arcsec$} & 1.8 & $(8.2\pm0.8)\cdot10^{-7}$ & $60.7\pm13.8$ & ...\\

    080804 & 319016 & 23:20:14 & \linebreakcell[t]{$21^{\rm h}54^{\rm m}40.12^{\rm s}$ \\ $-53\degr11\arcmin05.4\arcsec$} & 1.5 & $(3.6\pm0.2)\cdot10^{-6}$ & $34\pm16$ & 2.20\\

    080413A & 309096 & 02:54:19 & \linebreakcell[t]{$19^{\rm h}09^{\rm m}11.59^{\rm s}$ \\ $-27\degr40\arcmin41.1\arcsec$} & 2.3 & $(3.5\pm0.1)\cdot10^{-6}$ & $46\pm1$ & 2.43\\

    070805 & 287088 & 19:55:45 & \linebreakcell[t]{$16^{\rm h}20^{\rm m}13.8^{\rm s}$ \\ $-59\degr57\arcmin26 \arcsec$} & 90 & $(7.2\pm0.8)\cdot10^{-7}$ & $31.0\pm1.0$ & ...\\

    \hline
    110625A & 456073 & 21:08:28 & \linebreakcell[t]{$19^{\rm h}06^{\rm m}55.85^{\rm s}$ \\ $+06\degr45\arcmin19.2\arcsec$} & 2.1 & $(2.8\pm0.1)\cdot10^{-5}$ & $44.5\pm10.1$ & ...\\

    100418A & 419797 & 21:10:08 & \linebreakcell[t]{$17^{\rm h}05^{\rm m}27.18^{\rm s}$ \\ $+11\degr27\arcmin40.1\arcsec$} & 1.9 & $(3.4\pm0.5)\cdot10^{-7}$ & $7.0\pm1.0$ & 0.624\\

    091018 & 373172 & 20:48:19 & \linebreakcell[t]{$02^{\rm h}08^{\rm m}44.61^{\rm s}$ \\ $-57\degr32\arcmin53.7\arcsec$} & 1.7 & $(1.4\pm0.1)\cdot10^{-6}$& $4.4\pm0.6$ & 0.971\\

    090201 & 341749 & 17:47:02 & \linebreakcell[t]{$06^{\rm h}08^{\rm m}12.48^{\rm s}$ \\ $-46\degr35\arcmin25.6\arcsec$} & 1.4 & $(3.0\pm0.1)\cdot10^{-5}$ & $83\pm4$ & ...\\

    081221 & 337889 & 16:21:11 & \linebreakcell[t]{$01^{\rm h}03^{\rm m}10.20^{\rm s}$ \\ $-24\degr32\arcmin53.1\arcsec$} & 1.4 & $(1.81\pm0.03)\cdot10^{-5}$ & $34\pm1$ & 2.26\\

    070920B & 291728 & 21:04:32 & \linebreakcell[t]{$00^{\rm h}00^{\rm m}31.27^{\rm s}$ \\ $-34\degr51\arcmin10.2\arcsec$} & 8.0 & $(6.6\pm0.5)\cdot10^{-7}$ & $20.2\pm0.2$ & ...\\
    \hline
  \end{tabular}
}
\caption{Prompt (upper) and afterglow (lower part) GRB observations by H.E.S.S. between August 2007 and March 2012 after triggers from \emph{Swift}. The given fluence (15-150 keV) and T90 (15-350 keV) are measured by BAT, the position is typically the XRT position. An estimation for the redshift $z$ is not available for some GRBs listed.}
\label{tab:GRB_observations}
\end{center}
\end{table}

\begin{table}[p!]
\begin{center}
\footnotesize{
 \begin{tabular}{c|cccccc}
   & N$_{\rm{on}}$ & N$_{\rm{off}}$ & $\alpha$ & $N_{\rm excess}$ & $\sigma$\\
   \hline
   Total            & 46 & 427 & 0.12 & -4 & -0.6 \\
   First 300~s      &  8 &  39 & 0.13 &  3 &  1.2 \\
   $1^{\rm st}$ observation & 26 & 197 & 0.13 &  1 &  0.3 \\
   $2^{\rm nd}$ observation & 20 & 230 & 0.11 & -6 & -1.1 \\
   \hline
 \end{tabular}
 \caption{Preliminary results of the search of excess photons for the H.E.S.S. data on GRB~100621A. $N_{\rm{on}}$ is the number of gamma-ray candidates in the signal region around the GRB position and $N_{\rm{off}}$ the background estimate. When scaled by the normalisation factor $\alpha$ they yield the number of excess events $N_{\rm excess} = N_{\rm{on}}-\alpha N_{\rm{off}}$. The significance $\sigma$ is estimated using Eq. (17) of \cite{Li:1983fv}.}
 \label{tab:results_significance}
}
\end{center}
\end{table}

\end{document}